\documentclass[showpacs,showkeys,floatfix]{revtex4}
\usepackage[dvips]{graphicx}
\usepackage{epsfig}

\usepackage[cp1251]{inputenc}
\usepackage[english]{babel}

\usepackage{graphicx}
\begin{document}

\draft
\title{New mechanism for non-trivial intra-molecular vibrational dynamics.}

\author{V.A.Benderskii}
\affiliation {Institute of Chemical Physics Problems, 142432, Chernogolovka, Russia}

\author{E. I. Kats} \affiliation{Laue-Langevin Institute, F-38042,
Grenoble, France}
\affiliation{L. D. Landau Institute for Theoretical Physics, RAS, Moscow, Russia.}

\date{\today}

\begin{abstract}
 
We investigate the time evolution process of one selected (initially prepared by optical pumping) vibrational molecular state $S$, coupled to all 
other intra-molecular vibrational states $R$ of the same molecule,
and also to its environment $Q$. Molecular states forming the first reservoir $R$ are characterised by a discrete dense spectrum, whereas the 
environment reservoir $Q$ states form 
a continuous spectrum. Assuming the equidistant reservoir $R$ states we find the exact analytical solution of the quantum dynamic equations.
$S$ - $Q$ and $R$ - $Q$ couplings yield to spontaneous decay of the $S$ and $R$ states, whereas $S$ - $R$ exchange leads to recurrence
cycles and Loschmidt echo at frequencies of $S$ - $R$ transitions and double resonances at the interlevel reservoir $R$ transitions. 
Due to these couplings the system $S$ time evolution is not reduced to a simple exponential relaxation.
We predict various regimes of 
the system $S$ dynamics, ranging from exponential decay to irregular damped oscillations. 
Namely, we show that there are four possible dynamic regimes of the evolution:
(i) - independent of the environment $Q$ exponential decay suppressing backward $R$ -$S$ transitions,
(ii) Loschmidt echo regime, (iii) - incoherent dynamics with multicomponent Loschmidt echo, 
when the system state exchanges its energy with many states of the reservoir, (iv) - cycle mixing regime, when 
the long term system dynamics appear to be random.
We suggest applications of our results for interpretation of femtosecond vibration spectra of large molecules and nano-systems.

\end{abstract}


\maketitle

\section{Introduction}
\label{I}

A quasi-stationary state in quantum mechanics occurs as a result of overlapping (superposition)
of an initially localised (stationary) state with the states (i.e., wave functions) from continuous spectrum formed by either free states
of the same system (separated from the initial state by a potential barrier, as in the case
of  $\alpha $-decay), or by the system environment (see classical papers \cite{WW30}, \cite{SI39}).
Following this, common wisdom ascribes irreversible evolution of quasi-stationary states by coupling the states to a reservoir with continuous 
spectrum
\cite{GR93}, \cite{PP97}.
With a model reservoir formed by a sea of harmonic oscillators, this approach is at the heart of the theory of quantum dissipative systems
\cite{LC87}, \cite{WE99}, \cite{YS94}.

An opposite limit is considered in the theory of transition states which is widely used to treat various chemical dynamics problems
\cite{BM94}, \cite{BV00}. In this microscopic approach one has to choose properly a set of internal degrees of freedom forming a
so-called reaction path, and a small number of transverse degrees of freedom which are coupled to the longitudinal reaction coordinate.
This microscopic description is feasible in practice up to a few dozens of the transversal degrees of freedom. For larger
systems the microscopic approach becomes useless (since even with modern computer powers it is almost impossible to solve a system of 
dynamic equations with 
$10^4$
transversal degrees of freedom for all eigen states to restore multidimensional
potential energy surfaces). 
However many nano-systems with $10^2 - 10^4$ degrees of freedom,
interesting from their practical importance and the associated theoretical challenges, belong to an
intermediate case when both of the above mentioned
approaches (macroscopic theory of quantum dissipative systems and microscopic theory of chemical dynamics) do not work.
Evidently to cover these very complex phenomena, it is necessary as a first step at least to choose
an appropriate simple model.
One approach is to borrow concepts from other physical systems with dense discrete spectra, for example nuclei.
It was shown in \cite{MN08}, \cite{BT93}, \cite{KO01}, \cite{PW07}, \cite{ME68}, that
statistical description of such systems does not require any detailed information about its spectrum, but only
a few universal spectral characteristics (like interlevel spacing distribution function), determined by a random Hamiltonian matrix.
This approach is a very convenient tool to describe spectral chaos and many other global features of the behavior,
but it says almost nothing about quantum dynamics, in which we are interested in this paper. 
Our motivation is not a pure curiosity. As a matter of fact quantum dynamics of various systems (ranging from 
relatively small molecules in a pre-dissociation condition \cite{SH84}, \cite{SC94} up to large photochromic molecules and protein
complexes 
\cite{BM98}, \cite{JF02}, 
\cite{VB97}, \cite{BM98a}, \cite{HT03}, \cite{FE03}),
or molecules confined near interfaces \cite{BE02} (see also \cite{FE03})
is an active area of 
experimental research. Femtosecond spectroscopy data (which allows the study of the time evolution of one initially prepared
by the optical pumping state) manifest a variety of possible dynamic regimes including not only weakly damped more or less regular oscillations
but also very irregular long term behavior with a number of peaks corresponding to a partial recovery of the initial state
population. 

Seemingly irregular damped oscillation regimes
observed in
such systems  
 cannot
be explained theoretically in the frameworks of widely used
models with reservoirs possessing continuous spectra \cite{CL83}, \cite{LC87},
\cite{BM94}, \cite{YS94}, \cite{WE99}. Indeed in the case of a system coupled
to the continuous spectrum reservoir, 
only a smooth crossover between coherent
oscillations and an exponential decay is possible upon increasing of the coupling.
Nevertheless, as it was shown in the papers cited above, generic complex dynamics is observed
in the systems
with characteristic inter-level spacing of the order of $10 \, cm^{-1}$, when the measurements
are performed in the range of sub-picoseconds, or femtoseconds.
To explain these weakly damped oscillations 
semi-empirical models have been proposed  \cite{FE03}, \cite{RB93}, assuming more or less arbitrarily that the system interacts not only with
the environment (possessing continuous spectrum in the agreement with a common belief), but also with 
a few weakly damped discrete vibrational levels.
However, these models providing a possible mechanism for weakly damped oscillations, do not explain
irregular, random-like dynamic evolution. 
It is worth noting again that a generic feature of systems with such irregular behavior is
the existence of the dense but discrete vibrational spectra, about $10^2 - 10^3$ levels, and characteristic inter-level spacing 
of the order of $10 \, cm^{-1}$.
This generic feature is present in all systems with complex and irregular vibrational relaxation.
Motivated by these observations, our intent here is 
to propose a simple (but yet non-trivial) model
of a system coupled to a reservoir
with a discrete spectrum, and to examine
joint system-reservoir evolution, i.e., recurrence cycles, when the energy is flowing back from the reservoir to
the system, and to apply this model to intra-molecular vibrational
relaxation.

Vibrational relaxation in large molecules is one of the most relevant processes in chemical dynamics. The system time evolution
includes as its step intra-molecular energy transfer via transitions from a specially prepared and selected initial 
molecular state (system $S$) 
to a large finite number of distinct levels of the same molecules. This 
intra-molecular evolution should be supplemented,  of course, by the interaction of all these states that posseess a continuous 
spectrum environment, which ultimately yields to conventional exponential relaxation \cite{BJ68} - \cite{UM91}. 
To describe theoretically these radiationless transitions, usually one has to rely on approximate treatment of 
the system - reservoir interactions, e.g., on random phase approximation.
In this case the system time evolution is determined by the decay rate constant, which is calculated accordingly 
with the famous Fermi Golden Rule. Unfortunately such an approach
is valid only in the limit of a relatively weak system - reservoir coupling, 
when calculating perturbatively transition probability, one may neglect shift  
of the energy levels \cite{HO55}.

In our paper we develop another approach to analyse theoretically 
the problem of intra-molecular vibrational dynamics. Instead of a perturbative 
solution to the quantum dynamic equations for a system coupled to an arbitrary
reservoir, we find the exact solution for a simple model of a reservoir spectrum  (i.e., a set of final states). The model 
keeps the main essential feature of intra-molecular vibrational dynamics,
namely that realistic reservoir should possess discrete and dense spectrum of its states. 
Surprisingly enough, scanning the literature we could not find any paper treating theoretically such a model.
We do believe that the basic ideas inspiring our work can be applied to a large variety
of interesting nano-systems.
In the recent short publication \cite{BF07} such a model of a system coupled to a reservoir with dense discrete 
spectrum was proposed, and under assumptions put forward by Zwanzig \cite{ZW60}
(equidistant spectrum of the reservoir and system-reservoir coupling 
independent of reservoir state quantum numbers) its exact analytic solution was found.
Here we generalise this approach to rationalise intra-molecular vibrational relaxation
in terms of time evolution of quasi-stationary states.
The key feature, indispensable to provide energy exchange between the system and reservoir, is backward energy flow,
which is ignored in all models with continuous spectra of final states \cite{WE99}. 
These backward transitions (and corresponding recurrence cycles or Loschmidt echo phenomenon)
are the main new physical ingredients of our approach, and they are responsible for non-trivial time evolution.    
For a fairly broad class of
molecules, such a
non-trivial time evolution indeed was observed experimentally (see the monographs in \cite{ZE94}, \cite{TA92} and references therein)  
by modern femtosecond spectroscopy technique.
Our model seems to be adequate to rationalise these experimental observations, 
in a sense that we are not aiming at the best quantitative agreement of theoretical results 
and experimental data but at the identification, understanding
and explanation of important sources of non-exponential time evolution.

\section{Basic theoretical notions for a model with two reservoirs.}
\label{bas}

Our analysis is based on a simple but very general observation known from standard quantum mechanics.
For a selected unperturbed energy level $\epsilon _s^0$ (in what follows we will term the level as a system)
coupled to a discrete spectrum reservoir (with its unperturbed energy states $\{\epsilon _n^0\}$),
the Hamiltonian matrix contains besides the main diagonal, only one row and one line of non-zero matrix elements:
\begin{eqnarray} &&
\label{kats2}
\left |
\begin{array}{cccccccc}
... &... &... & ... & .... & ... &... &...\\
... & \epsilon _1 & 0 & 0 & C_1 & 0 & ... \\
... & 0 & \epsilon _2 & 0 & C_2  & 0 & ...\\
... & 0 & 0 & \epsilon _3 &  C_3 & 0 & ...\\
... & C_1 & C_2 & C_3 & \epsilon _4 & C_4 & ... \\
... & 0 & 0 & 0 & C_4 & \epsilon _5 & ... \\
... & 0 & 0 & 0 & C_5 &  0 & ...\\
... & ... & ... & ... & ... & ... &...&...
\end{array}
\right | \, .
\end{eqnarray}
Therefore the corresponding secular equation (its roots determine the coupled system-reservoir eigenvalues)
has the following deceptively simple form \cite{BF07}, \cite{BK09}
\begin{equation}\label{b1}
F(\epsilon ) = \epsilon - \sum _{n} \frac{C_n^2}{\epsilon -\epsilon _ n^0} = 0
\, ,
\end{equation}
where we count the energy levels from $\epsilon _s^0$, and to get such a compact form for the function $F(\epsilon )$
we use the orthogonal basis of the reservoir states, i.e., all matrix elements between the reservoir states are zero.
With the same approach we can include all ingredients of radiationless molecular transitions into a simple model, 
namely
\begin{itemize}
\item 
system $S$ which is selected molecular level
\item
discrete dense spectrum $\{\epsilon _n^0\}$ reservoir $R$, which includes all other than $S$ molecular states
\item
$S - R$ coupling characterizing by the coupling matrix elements $\{C_n\}$
\item
second reservoir (environment) $Q$ with continuous spectrum, which leads to a decay of the system $S$ (with its rate
constant $\Gamma _s $) and of the reservoir $R$ states (with corresponding constants $\{\Gamma _n\}$).
\end{itemize} 
One can derive formally an exact secular equation similar to the (\ref{b1}) for the system $S$ and reservoir $R$ eigenvalues,
which additionally includes the environmental level broadening (due to $S - Q$, and $R - Q$ couplings).
The corresponding determinant contains (besides the diagonal matrix elements $E + i \Gamma _s$, and $\{E - \epsilon _n^0 + i \Gamma _n\}$)
only one non-zero row and one line with non-zero $S - R$ coupling matrix elements. 
The corresponding secular equation  
reads as
\begin{eqnarray}
\label{bk1}
F(E) \equiv E + i \Gamma _s - \sum _{n}\frac{C_n^2}{E - \epsilon _n^0 + i \Gamma _n} = 0
\, .
\end{eqnarray}
Of course for arbitrary functions $C_n$, $\epsilon _n^0$, and $\Gamma _n$, the sum in the Eq. (\ref{bk1}) cannot 
be calculated analytically. To proceed further on we are following the Zwanzig idea \cite{ZW60}
(already broadly used in the theory of radiationless transitions \cite{BJ68} - \cite{UM91}, and time resolved 
spectroscopy \cite{TA92}, \cite{MH86}, \cite{KM88}) assuming that
\begin{eqnarray}
\label{bk2}
\epsilon _n^0 = n\, ;\, C_n = C\, ; \, \Gamma _n = \Gamma
\, ,
\end{eqnarray} 
where as above, we count all energy levels from the system $S$ energy unperturbed by its coupling to the reservoir levels, 
and the unperturbed reservoir $R$ interlevel spacing is chosen as 
the energy unit. It is worth noting also that the approximation of equidistant $R$ reservoir spectrum
is not a completely artificial one. Indeed, for any system with a sufficiently large finite
number of degrees of freedom, mutual level repulsion unavoidably favours to more or less equidistant interlevel spacing.
A more realistic model will not affect our qualitative
conclusions, and transparency is worth a few simplifications. 
If the Zwanzig model assumptions are granted, the secular equation (\ref{bk1}) can be written in the following compact and easily solvable form
\begin{eqnarray}
\label{bk3}
F(E) = E + i \Gamma _s - \pi C^2\cot [\pi (E + i \Gamma )] = 0
\, .
\end{eqnarray}
System $S$ and reservoir $R$ energy states can be regarded as quasi-stationary ones if \cite{ZE61}
\begin{eqnarray}
\label{bk4}
\Gamma _s \ll 1\, ; \, \Gamma \ll 1
\, ,
\end{eqnarray}
then complex solutions to the secular equation (\ref{bk3}) are
\begin{eqnarray}
\label{bk5}
E_n = \epsilon _n - i \gamma _n
\, ,
\end{eqnarray}
where
\begin{eqnarray}
\label{bk6}
\epsilon _n = \pi C^2 \cot (\pi \epsilon _n) + (\Gamma _s - \Gamma )^2
\frac{[\epsilon _n \cot (\pi \epsilon _n) + \pi C^2]\cot (\pi \epsilon _n)}{[\epsilon _n \cot (\pi \epsilon _n) + \pi C^2 + (1/\pi )]^2}
\, ,
\end{eqnarray}
and
\begin{eqnarray}
\label{bk7}
\gamma _n = \Gamma + (\Gamma _s - \Gamma )\frac{1}{\epsilon _n \cot(\pi \epsilon _n) + \pi C^2 + (1/\pi )}
\, .
\end{eqnarray}
Note that the shift of the zero level energy corresponds to a change of variables in a rotating frame,
and it does not affect system dynamics in which we are interested in this work.
By a simple inspection of the Eqs. (\ref{bk6}) - (\ref{bk7}) we conclude that $S - Q$ and $R - Q$ couplings
displace the system $S$ and reservoir $R$ levels into the lower complex semiplane with energy dependent decay rates.
$S - Q$, $R - Q$, and $S - R$ interactions yield to variations of all energy levels. 
It is easy to see that the energy level shifts ''$\epsilon _n - n$'' are mainly due to $S - R$ coupling. 
It means that independent of the coupling strength $C$ there is one energy level in each interval $[n \, ,\, n+1]$.

\section{System amplitude}
\label{amp}

To analyse the system time evolution we have to solve the time dependent equations of motion (i.e., the corresponding 
Heisenberg equations of motion) for the system state amplitude $a_s$. 
Our Hamiltonian model describing the system $S$ and the discrete spectrum reservoir $R$ (while producing decay of the
quasi-stationary states couplings with the continuous spectrum reservoir $Q$ are taking into account by introducing
the decay rates $\Gamma _s$ and $\Gamma _n$ into the equations)
\begin{eqnarray}
\label{jl1}
H = \epsilon _s^0 b_s^+ b_s + \sum _{n} \epsilon _n^0 b_n^+ b_n + \sum _nC_n(b_s^+ b_n + b_s b_n^+)
\, ,
\end{eqnarray}
where $b_s^+$, $b_n^+$ are corresponding creation operators.
Time dependent wave functions $\Psi _s(t)$ of the Hamiltonian (\ref{jl1}) can be expanded over the unperturbed (uncoupled)
eigenfunctions of the system $\Phi _s$ and of the reservoir states $\Phi _n$
\begin{eqnarray}
\label{jl21}
\Psi _s (t) = a_s(t) \Phi _s + \sum _{n} a_n(t)\Phi _n
\,
\end{eqnarray}
with time dependent amplitudes $a_s(t)$, $a_n(t)$.
These time dependent amplitudes  satisfy the corresponding Heisenberg equations of motion.
Combining everything together 
we end up with a general case (i.e., not applying immediately the Zwanzig approximation), the equations read
(in units with $\hbar =1$, and dots steam for time derivatives) \cite{BF07}, \cite{BK09}
\begin{eqnarray}
\label{bk8}
i {\dot {a}_s} = - i \Gamma _s a_s + \sum _{n} C_n a_n\, ;\, 
i {\dot {a}_n} = (\epsilon _n^0 - i \Gamma _n) a_n + C_n a_s
\, .
\end{eqnarray}
The solution to the Eq. (\ref{bk8}) supplemented by the initial conditions
\begin{eqnarray}
\label{bk9}
a_s(0) = 1\, ;\, a_n(0) =0 
\, ,
\end{eqnarray}
can be formally found as
\begin{eqnarray}
\label{bk10}
a_s(t) = \frac{1}{2 \pi i}\int _{-\infty }^{\infty } d E \exp (-i E t) 
\left (E + i \Gamma _s - \sum _{n} \frac{C_n^2}{E + i \Gamma _n - \epsilon _n^0} \right )^{-1}
\, .
\end{eqnarray} 
The amplitude can also be expressed as a sum of residues in the poles of the integrand in the Eq. (\ref{bk10}) which are the roots of the secular 
equation (\ref{bk1})
\begin{eqnarray}
\label{bk11}
a_s(t) = \sum _{n}\left .\frac{\exp (-i E t)}{dF/dE}\right |_{F(E) = 0}
\, .
\end{eqnarray} 
Let us apply now the Zwanzig approximation (\ref{bk2}). In this case the formal solution for the system amplitude (\ref{bk11}) can be written down in 
the 
explicit analytical form
\begin{eqnarray}
\label{bk12}
a_s(t) = \sum _{n = -\infty }^{n = \infty }\frac{\exp (-i E_n t)}{1 + \pi ^2 C^2 + C^{-2}(E_n + i \Gamma _s)^2}
= 2 \sum _{n=0}^{n=\infty }\exp (- \gamma _n t)[B_n \cos (\epsilon _n t) + S_n \sin (\epsilon _n t)]
\, ,
\end{eqnarray}  
where
\begin{eqnarray}
\label{bk13}
B_n = B[\epsilon _n^2 - (\Gamma _s - \gamma _n)^2 + C^2(1+\pi C^2)]\, ; \,
\, ,
\end{eqnarray} 
and
\begin{eqnarray}
\label{bk133}
S_n = 2B\epsilon _n(\Gamma _s - \gamma _n)
\, .
\end{eqnarray}
The coefficient $B$ in these formulas reads as
\begin{eqnarray}
\label{bk14}
B = C^2[(\epsilon _n^2 -(\Gamma _s - \gamma _n)^2 + C^2(1 + \pi ^2 C)]^2 + 4\epsilon _n^2(\Gamma _s - \gamma _n)^2]^{-1}
\, .
\end{eqnarray} 
The above expressions (\ref{bk12}) - (\ref{bk14}) give the exact solution to describe quantum dynamics of the system $S$ coupled to 
the Zwanzig reservoir $R$ and continuous spectrum environment reservoir $Q$.

\section{Recurrence cycle amplitude representation}
\label{par}
Presented in the previous section exact results are not very useful in practical terms, because of
a poor convergence of Fourier series entering these expressions. Luckily one can transform the results 
into a form more convenient for further calculations (both analytical and numerical) by
replacing the Fourier expansion with the expansion over partial recurrence cycle amplitudes. 
The method of how to do it, developed in the papers \cite{BF07}, \cite{BK08}, \cite{BK09}, is based on the Poisson summation formula, 
(see e.g., \cite{MF53}). To transform $a_s(t)$ as it is given in the (\ref{bk12}), we introduce the following formal identity for 
the $\delta $ function
\begin{eqnarray}
\label{bk15}
\delta (E(x)) = \sum _{n}\left |\frac{d E}{d x}\right |^{-1} \delta (x - n)
\, .
\end{eqnarray}
The Eq. (\ref{bk15}) allows us to replace the discrete variables $n$ and $E_n$ by their continuous counterparts $x$ and $E(x)$ with $E(n) \equiv E_n$.
Thus we end up with
\begin{eqnarray}
\label{bk16}
a_s(t) = \sum _{n}\left (\frac{dF}{dx}\right )^{-1}\exp (-iEt)\delta (E - E_n)=
\end{eqnarray}
\begin{eqnarray}
\nonumber
\sum _{k=-\infty }^{\infty }
\int _{-\infty }^{\infty }dE\left (\frac{dF}{dE}\right )^{-1}\left (\frac{dE}{dx}\right )^{-1}\exp [-iE(t - 2k \pi) + i2k\pi (E - x)]
. 
\end{eqnarray}
The expression (\ref{bk16}) represents system evolution as a sum over recurrence cycles
\begin{eqnarray}
\label{bk17}
a_s(t)=\sum _{k = -\infty }^{\infty } a_s^{(k)}(t - 2 k \pi )
\, ,
\end{eqnarray}
where the partial cycle $k$ amplitude (for Zwanzig reservoir $R$) reads as
\begin{eqnarray}
\label{bk18}
a_s^{(k)}(t-2 k \pi ) = C^2 \exp (-2 k \pi \Gamma )\int _{-\infty }^{\infty }d E \frac{(E + i \Gamma _s - i \pi C^2)^{k-1}}{(E + i \Gamma _s + 
i \pi C^2)^{k+1}}\exp [- iE(t - 2k \pi)]
\, .
\end{eqnarray}
In the initial cycle $k=0$, there are two poles $E_\pm ^* = - i(\Gamma _s \pm \pi C^2)$ in the integrand 
for $a_s^{(0)}$, and for $t > 0$ only the pole $E^*_+$ contributes, yielding to the exponential relaxation law for the $a_s^{(0)}(t)$
\begin{eqnarray}
\label{bk19}
a_s^{(0)}(t) = \exp [- (\Gamma _s + \pi C^2)]
\, .
\end{eqnarray}
We see that $ S - R$ coupling gives an additional (with respect to the environment $Q$) decay channel
with its rate 
\begin{eqnarray}
\label{new1}
\Gamma _R = \pi C^2
\, .
\end{eqnarray}
In a higher order recurrence cycles $ k \geq 1$, the pole $E^*_-$ disappears and the residue in the $(k+1)$-th order pole $E^*_+$ leads to
\begin{eqnarray}
\label{bk191}
a_s^{(k)}(t-2k \pi ) = \exp [- 2k\pi \Gamma ] \exp [- \Gamma _s (t - 2 k \pi )] a^{(k)}_{s0}
\, ,
\end{eqnarray}
where for the sake of compactness
\begin{eqnarray}
\label{bk192}
a_{s0}^{(k)}(t) = - \frac{2\Gamma _R(t - 2 k\pi )}{k} L^1_{k-1} [2\Gamma _R(t - 2k \pi )]\Theta (t- 2k \pi )
\, .
\end{eqnarray}
Here $ k \geq 1$, $L_{k-1}^1$ is adjoint Laguerre polynomial \cite{BE53}, and $\Theta (z)$ is a step function
defined as 
$\Theta (z<0) = 0$, and
$\Theta (z>0) = 1$. Since the pole $E_+^*$ exists only for $t \geq 2 k \pi $ in the representation of the recurrence cycle partial 
amplitudes 
for $a_s(t)$ only the cycles with $k$ ranging from 0 to $[t/2\pi ]$ ($[x]$ stands for the integer part of $x$) contribute into the sum entering
equations (\ref{bk16}) - (\ref{bk17}).

Similarly the amplitudes for the reservoir $R$ states can be written as
\begin{eqnarray}
\label{bk193}
a_n(t) = \sum _{k = 0}^{[t/2\pi ]} a_n^{(k)}(t - 2 k \pi )
\, ,
\end{eqnarray}
where for $k=0$
\begin{eqnarray}
\label{bk194}
a_n^{(0)}(t) = \frac{C}{n + i (\Gamma _s + \Gamma _R - \Gamma )}(\exp [-(\Gamma _s + \Gamma _R) t] - \exp [-(i n + \Gamma ) t])
\, ,
\end{eqnarray}
and for $k \geq 1$
\begin{eqnarray}
\label{bk195}
a_n^{(k)}(t - 2 k \pi ) = - i C \exp [-(in + \Gamma )(t - 2 k \pi )]\int _{0}^{(t - 2 k \pi )} dt^\prime \exp ((i n + \Gamma )t^\prime ) 
a_s^{(k)}(t^\prime )
\, .
\end{eqnarray}
It is worth noting that although transformation from the Fourier representation into the partial amplitude 
representation mathematically is merely a change of variables, there is some physics behind it. 
By inspection of  both representations one can see that the Fourier expansion coefficients are determined by the real eigenvalues, 
while the 
partial amplitudes depend mainly on imaginary pole displacements (which in their own turn are related to the $S - R$ 
coupling or decay rate $\Gamma _R$). This is not a minor technical issue but an essential physical difference, 
because in fact it means that the partial recurrence cycle amplitudes take into account effects of all $R$ states. 
Physics occurs as a result of it. Indeed quantum interference of transitions from different $R$ 
states suppresses effectively 
back transitions from the $R$ states into the system $S$, 
and yields to a new decay channel. Its decay rate is determined by the transition probability 
into all $R$ states. Back transitions from the $R$ states into the system 
lead also to the spontaneous Loschmidt echo at each recurrence cycle. Instead of an individual level-to-level transition 
picture (satisfying standard detailed balance relations) we have to deal with cooperative 
transitions from $S$ into all $R$ states. $\Gamma _R$ is the rate of these transitions, and for 
the backward transitions from $R$ to $S$ the effective rate constant is of the order of unity, because 
the reservoir $R$ should ''wait'' for synchronisation of its states to have backward $R$ to $S$ transition.

This non-trivial behavior manifests itself in a fine structure of the Loschmidt echo signals. 
Utilising known properties of the Laguerre polynomials \cite{BE53} we arrive at the conclusion similar to that found in 
\cite{BF07}, \cite{BK08}, \cite{BK09} for a single reservoir Zwanzig model. 
Namely, that in the $k$-th recurrence cycle, the Loschmidt echo signal has $k$ components, $I_{k l}$ with $l = 1,.....,k$. 
The component $I_{kk}$ is the most intense one. Because different $R$ states exchange 
their populations with the system $S$ not at the same time, the echo signal is broadened. 
The full width of the signal in the cycle $k$ is about $4 k$. In its own turn when $k$ grows,
the Loschmidt echo signals for the neighboring cycles start to overlap, if $ k > k_c = \pi ^2 C^2$, and for $k \gg k_c$
time evolution looks as irregular (quasi-stochastic) damped oscillations.

It is also interesting to compare the formally exact presentation of $a_s(t)$ in terms of the partial cycle amplitudes
(\ref{bk17}) - (\ref{bk18}) with its approximate treatment according to the Golden Fermi Rule. The equations of motion
(\ref{bk8}) written in the equivalent integral form
can be rewritten as the quantum Langevin equation
for kicked harmonic oscillations with friction 
\begin{eqnarray}
\label{bk20}
{\dot {a}}_s(t) = - \Gamma _s a_s - \int _{0}^{t} d t^\prime a_s(t^\prime )\sum _{n} C_n^2 \exp [-i(\epsilon _n^0 - i\Gamma _n)(t - t^\prime )]
\, .
\end{eqnarray}
The terms entering this equation have transparent physical meaning:
the resonance transition to the level $n=0$ of the reservoir leads to coherent oscillations with frequency $\propto C$, whereas
non-resonance ($n \neq 0$) transitions produce effective friction and exciting force.
However, strictly speaking, there is no stochasticity in the equation, and we have fully deterministic energy levels and therefore 
time evolution.

In the Golden Fermi Rule approximation we replace
\begin{eqnarray}
\label{bk21}
\sum _{n} \to \int _{-\infty }^{\infty }\rho (\epsilon ) d \epsilon
\, ,
\end{eqnarray}
with $\rho (\epsilon )$ being unperturbed (by interactions) density of states.
Furthermore, if $C_n^2(\epsilon _n^0)\rho (\epsilon _0)$ is a slow varying (in comparison to exponential) function, 
the integrand in the (\ref{bk20}) - (\ref{bk21}) can be approximated by the mean value of this function. Thus, according to the 
Golden Fermi Rule
\begin{eqnarray}
\label{bk22}
{\dot {a}}_s(t) = - (\Gamma _s + \Gamma _{GFR})
\, ,
\end{eqnarray}
where $\Gamma _{GFR} = 2 \pi \langle C^2(\epsilon ) \rho (\epsilon )\rangle $, and therefore in this approximation 
the memory about all recurrence cycles is lost. For a single reservoir Zwanzig model \cite{BK08} 
this approximation corresponds to keeping only one term $k=0$ in the Poisson summation formula.

\section{Time evolution regimes}
\label{TE}
The partial cycle amplitudes $a_s^{(k)}$ are represented 
according to the expressions (\ref{bk191}), (\ref{bk192}) as a product of three factors. The first factor is related to the $R - Q$ coupling. 
Physically it 
describes the Loschmidt echo signal reduction due to spontaneous decay of the $R$ states in the previous (with respect to the cycle $k$ under 
consideration) cycles. In the strong coupling limit we are interested in, i.e., for $\Gamma _R \gg 1$, 
this factor does not depend on $\Gamma _s$. Indeed in this limit in the major part of the recurrence 
cycle only the reservoir $R$ states are populated, whereas the system population is negligibly small. 
The second factor in the Eqs. (\ref{bk191}), (\ref{bk192}) is related to the system decay in the same cycle $k$. 
Naturally this factor does not depend on the decay rate of the reservoir $R$ states. 
Finally the third factor in the (\ref{bk191}), (\ref{bk192}) is related to the decay rate the system at $\Gamma _s = \Gamma = 0$.

These relations (\ref{bk191}) - (\ref{bk192}) and their interpretation are our main results in this paper. 
Analysing the relations we end up with the following classification of the possible time evolution regimes. Namely
\begin{itemize}
\item
Exponential decay in the initial cycle $k=0$, when the Loschmidt echo intensity is exponentially small
\begin{eqnarray}
\label{bk23}
\Gamma \gg \frac{1}{2\pi }
\, .
\end{eqnarray}
If $\Gamma _R \gg \Gamma _s$ then the decay rate $\Gamma _R + \Gamma _s \simeq \Gamma _R$ does not depend on the reservoir $Q$.
Physically it means that in this limit the rate of radiationless
intra-molecular transitions is independent of the molecular environment $Q$. The states of the $R$ reservoir with
$|n| \leq \Gamma _R$ gives the main contributions into the $S$ state decay rate (dominated in this limit by the $S$ - $R$
exchange). These reservoir states populations monotonously increase at $\Gamma _R t \leq 1$ and achieve maximal values at $T^*
\simeq \Gamma _R^{-1}\ln (\Gamma _R/\Gamma )$. Then the state $n$ population decreases as
\begin{eqnarray}
\label{bk233}
|a_n(t)|^2 = \frac{\Gamma _R}{\pi (n^2 + (\Gamma _R + \gamma _s - \Gamma )^2)}\left (\exp (- \Gamma t) - \exp (-(\Gamma _r + \Gamma _s)t)
\right )
\, .
\end{eqnarray}
\item
Regular Loschmidt echo regime, where stochastic-like behavior is not achievable, because the echo intensity is already negligibly
small when the cycles start to mix
\begin{eqnarray}
\label{bk24}
\frac{1}{2 \pi } > \Gamma > \frac{1}{2\pi k_c}\, ; \, \Gamma _R \geq 1 > \Gamma _s
\, .
\end{eqnarray}
In this regime we get Loschmidt echo phenomenon which occurs due to synchronisation of the backward (from the reservoir
$R$ into the system $S$) transitions. Furthermore when the state $n$ of the reservoir is de-populated completely,
its complex amplitude $a_n(t)$ acquires the phase shift $\pi $. 
This leads to resonance behavior at the double reservoir $R$ interlevel
transition frequency. In turn these double transitions $n \to s \to n^\prime $ effectively redistribute the population 
between the reservoir $R$ states. Besides the phase shift $\pi $ yields to $a_s(t)$ sign alternation
between even and odd cycle numbers (see the Fig. 1). In odd cycle numbers the energy (or population) flows
from the reservoir $R$ into the system, whereas in even cycle numbers it flows from the system into the reservoir states.
\item
Quasi-stochastic time evolution when
\begin{eqnarray}
\label{bk25}
\frac{1}{2\pi } > \Gamma > \frac{1}{2\pi k_c}\, ; \, 1 > \Gamma _R > \Gamma _s 
\, ,
\end{eqnarray}
when the system time evolution is described by incoherent dynamics with multicomponent Loschmidt echo.
In this case the system exchanges its energy with 
many states of the reservoir.
\item
Cycle mixing regime
\begin{eqnarray}
\label{bk244}
\Gamma \ll \frac{1}{2\pi k_c}\, ; \, \Gamma _R \gg 1 \gg \Gamma _s
\, .
\end{eqnarray}
In this case the cycles are strongly overlapped, the system amplitude oscillates in a shorter
and shorter time scale, because
the number of zeros of the $a_s(t)^{(k)}$ in the cycle $k$ increases proportionally with $k$.
Therefore to keep fully deterministic system description at $t \to \infty $,
one has to know the system evolution at shorter and shorter time intervals.
This phenomenon leads to effective
loss of deterministic dynamics
at arbitrary small time coarse-graining, related physically to uncertainties
of the system or of the reservoir, for example.
\end{itemize}
We see from these conditions that the Loschmidt echo occurs depending mainly on the decay rate of the $R$ states,
whereas stochastic like behavior occurs depending on the product of $\Gamma \cdot \Gamma _R$, which characterises the energy (or level population) 
flow 
from the system into the environment $Q$ through the reservoir $R$ states as intermediate ones. 
We illustrate these findings in the Fig. 1 which shows how system level $\epsilon _s$ population ($ \propto |a_s(t)|^2$) depends on the coupling to 
the reservoir $R$ states.
\begin{figure}[htbp]
{\centerline{\psfig{file=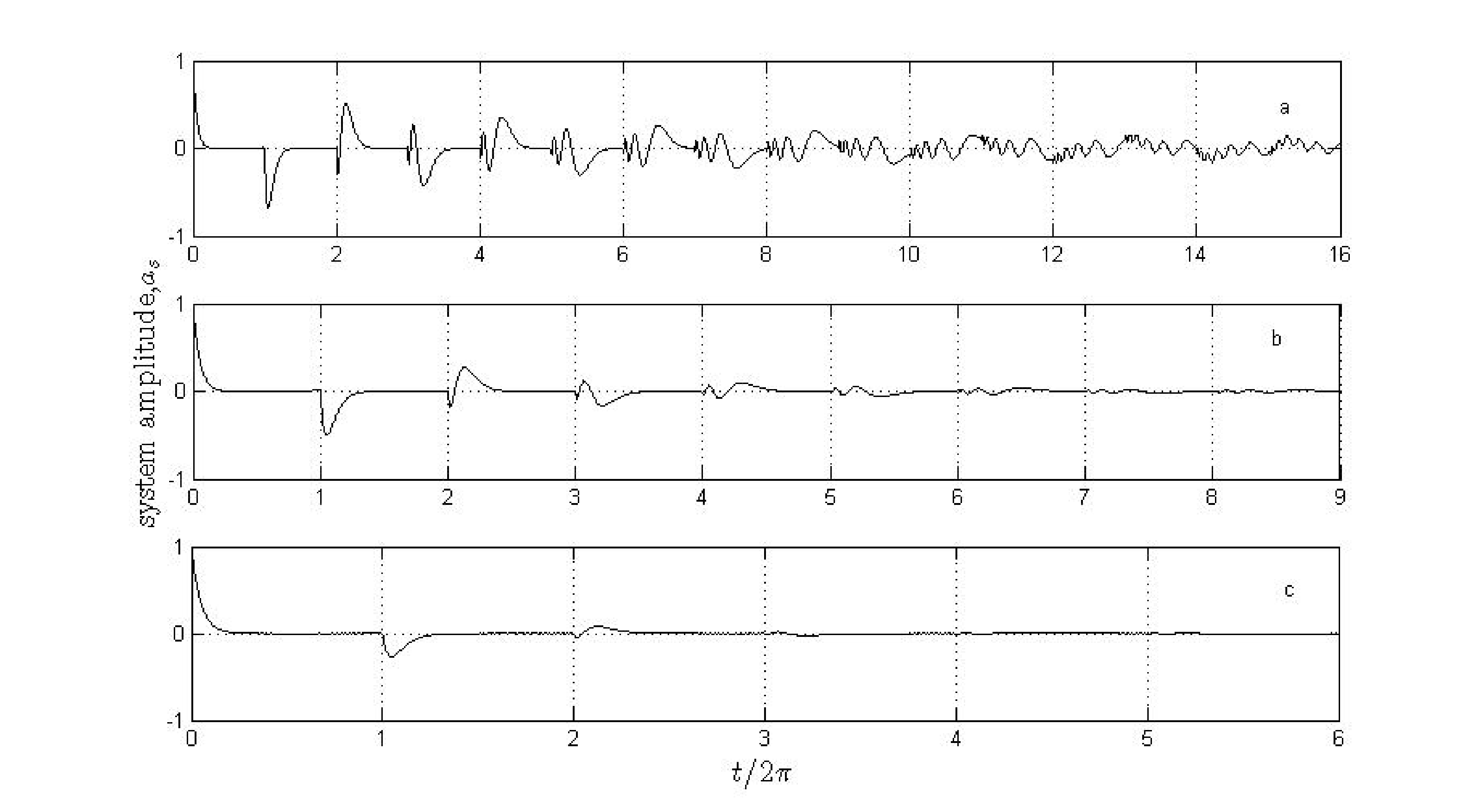,width=8cm} }}
\caption{System population time evolution for the model parameters: $C^2 =1$, $\Gamma _s =0.2$.
$\Gamma $ shows how strong intramolecular vibrations are coupled to environment:
(a) $\Gamma = 0.01 $; (b) $\Gamma = 0.05 $; (c) $\Gamma = 0.15 $.
Time is measured in the recurrence cycle period units $2\pi /\Omega $, where $\Omega $ is the interlevel spacing in the reservoir $R$.
}
\label{f1}
\end{figure}
The $S - Q$ coupling affects also the shape of the Loschmidt echo signal, 
suppressing upon $\Gamma _s$ growing the most intensive echo component $I_{kk}$.
Our results demonstrate the main advantage of the partial cycle amplitude 
representation in comparison with the standard Fourier representation for the $a_s(t)$ evolution. 
The latter one depends on the complex manner of interplay of all three relevant coupling 
parameters $ S - Q$, $R - Q$, and $S - R$, whereas in the former one these three interactions are factorised.

\subsection{Averaged over cycle population of state}
\label{av}
The system $S$ population averaged over a cycle $k$ is determined naturally as
\begin{eqnarray}
\label{bk26}
\langle |a_s^{(k)}|^2\rangle _k \equiv \int _{0}^{2\pi } dt^\prime |a_s^{(k)}(t^\prime )|^2
\, .
\end{eqnarray}
For non-overlapping cycles the upper integration limit in the (\ref{bk26}) can be moved to infinity (since the integrand is exponentially small for 
$t^\prime > 2 \pi $). Replacing $a_s^{(k)}(t^\prime )$ in the (\ref{bk26}) by its Laguerre polynomial explicit form (\ref{bk192}), and utilising
known \cite{BE53} integrals with Laguerre polynomials we find
\begin{eqnarray}
\label{bk27}
\langle |a_s^{(k)}|^2\rangle _k \simeq \Gamma _R^{-1}\exp (-4\pi \Gamma )(1+\zeta )^{-3}\frac{F(-k+1,3/2,-k+3/2,(1-\zeta)/(1+\zeta))^3)}
{F(-k+1,3/2,-k+3/2,1)}
\, ,
\end{eqnarray}
where $F(a,b,c,z)$ is a hyper-geometric function \cite{BE53}, and $\zeta \equiv \Gamma _s/\Gamma _R$.
When $\zeta \ll 1$, the (\ref{bk27}) becomes $\zeta $-independent, and the averaged over cycle system state population decays with
the $R - Q$ transitions rate constant $4\pi \Gamma $. Let us stress again the message we got from the comparison of the expression (\ref{bk19}) for 
the initial cycle 
decay rate and the expression (\ref{bk27}) for long-time evolution: the former is determined by the $S - Q$ transitions, whereas the latter one by the 
$R - Q$
transitions.

\subsection{Notes on the coarse grained spectrum}
\label{ab}
As it was pointed out above, when $k$ (cycle number) is growing, the number of the fine structure components of the Loschmidt echo signals
grows proportionally to $k$. As a result of increasing $k$, the echo shape becomes more and more complex. To detect the 
Loschmidt echo
signal accurately one has to use a detector with better and better 
resolution because the signal shape upon $k$ growing includes more and more components.
The same can be formulated that for a given detector resolution, there is always a certain cycle number, when
the accuracy of the detector is not sufficient to resolve the Loschmidt echo signal.
Therefore the finite accuracy of any detector device yields to the fact that one may not restore the energy spectrum by measuring the time 
evolution
of the system amplitudes $a_s(t)$. Although mathematically speaking to restore the spectrum it is enough to perform inverse Fourier transform of the 
amplitude, in 
practical terms it is not possible, since the amplitude cannot be measured with the required accuracy for high enough cycle number.
With a finite detector resolution we are able to perform only truncated Fourier transformation up to a certain limiting recurrence cycle number  
threshold 
$K_m$. 
We illustrate this message on a reduced accuracy of truncated Fourier transformation in the Fig. 2. 
Note that noise from the measurement device also restricts a spectrum part which can be reproduced from asymptotic
behavior of $a_s(t)$. The Fig. 2c shows spectrum distortion induced by external white noise.
\begin{figure}[htbp]
{\centerline{\psfig{file=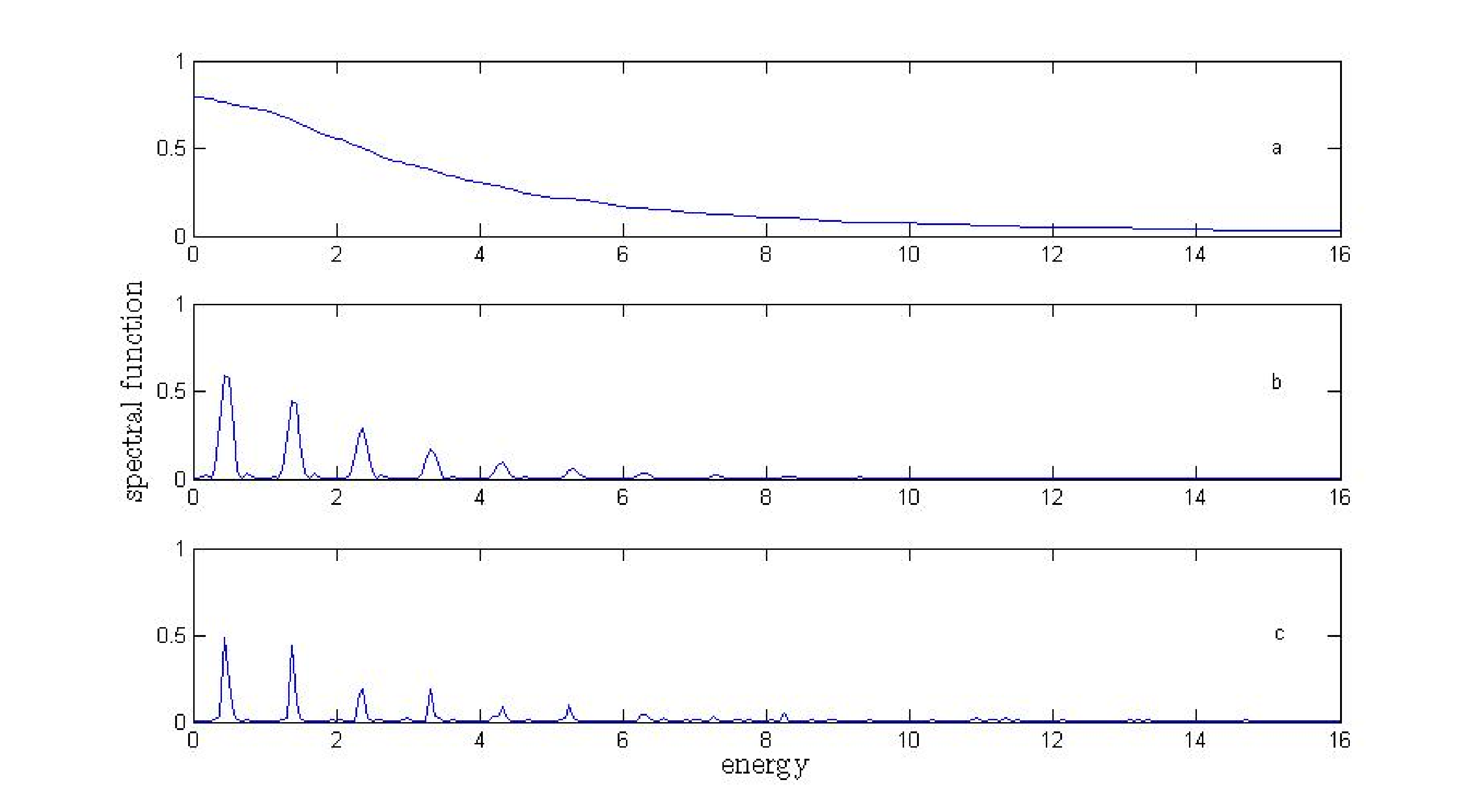,width=8cm} }}
\caption{Adsorption spectrum for the reservoir $R$ calculated by truncated Fourier transformation of the amplitudes
for the model parameters $\Gamma = \Gamma _s = 0.1$, $C^2 = 1$, and the limiting by measurement precision recurrence cycle number $K_m$
is:
(a) $K_m=1$; (b) $K_m = 5$; (c) $K_m = 20$. In the Fig. 2c white noise produced by a measurement device (signal/noise ratio 10/1)
is also added.
}
\label{f2}
\end{figure}
Namely, we have calculated the $R$
-reservoir absorption spectrum for the truncated at $K_m$ Fourier transformation of the amplitudes. We see that coarse graining produced
by  spectral measurement device, reduces the reproducible size of the spectrum, and distorts strongly the line shape.
In the Fig. 2a ($K_m$ is sufficiently small) no traces from the recurrence cycles are visible, and we get Fermi Golden Rule
exponential decay.
For the larger values of the $K_m$ (Fig. 2b) there are clear indications for adsorption peaks corresponding to
the Loschmidt echo signals in $a_s(t)$. However, adding external white noise (Fig. 2c) we see that the adsorption peaks are reduced even for the 
bigger (than in the Fig. 2b) values of $K_m$.
In its spirit the described mechanism is in the heart of quantum mechanical uncertainty produced by the interaction
between quantum system and classical measurement device. It is considered as a route to quantum chaos and irreversibility phenomena
\cite{ZU82}, \cite{GR93}.

\section{Conclusion}
\label{con}
To summarise,
in this paper we investigated quantum
dynamics of a single selected state of a small system. This selected state is coupled to other discrete dense states of the same system,
and all the states are coupled also to continuous spectrum environment.
This publication  represents a substantial extension of the note \cite{BF07}, where only a single equidistant
reservoir and constant-coupling Zwanzig model have been studied. Here we provide a more complete account and investigation of phenomena only
briefly addressed in \cite{BF07}, and generalise the Zwanzig model to include 
quasi-stationary states and coupling to
the second continuous spectrum reservoir - environment
$Q$. 
We show that there are possible four dynamic regimes of the evolution:
\begin{itemize}
\item
(i) - independent of the environment $Q$ exponential decay suppressing backward $R$ -$S$ transitions
\item
(ii)
Loschmidt echo phenomenon occurring not only due to almost coherent oscillations governed by 
transitions from the system to the resonance reservoir $R$ state, but also due to double transitions $n\, \to s\, \to n^\prime $
at the double reservoir $R$ interlevel
transition frequency
\item
(iii) - incoherent dynamics with multicomponent Loschmidt echo, when the system is exchanged its energy with many states of the reservoir
\item
(iv) - cycle mixing regime, when due to unavoidable coarse graining 
in any real system
of time or energy measurements, or initial condition uncertainty,
the system loses invariance with respect to time inversion.
In such conditions dynamic evolution of the system cannot be determined uniquely from the spectrum, and
in this sense long term system dynamics appear to be random.
\end{itemize}
The quantum dynamics 
of the selected level demonstrates non-trivial fine structures of the recurrence cycles
(Loschmidt echo) and cycle mixing leading eventually to irregular, chaotic-like long term evolution.
Our results illustrate non-ergodic dynamics of such a system, i.e., system population (or its energy)
is not equally distributed over all system states but in certain time intervals it is concentrated in a few levels.
The generalised Zwanzig model investigated in this paper reflects the spirit of minimalist approaches, in that it is simple yet based on a
physical principle.
The results presented here are probably less notable in terms of technological applications of nano-systems,
than
for the progress they could generate in our understanding of their complicated vibrational spectra.

Our consideration
yields quite reasonable qualitative description
of a variety of vibrational relaxation regimes and mode selectivity
observed in experiments, and the model under investigation appears to be the simplest one demonstrating that relatively small
variation of the coupling
enables us to change qualitatively the dynamic behavior.
One of the main difficulties in a quantitative comparison of the results of our simplified model with
specific experimental measurements or elaborated numerical simulations is the availability of an accurate connection
between experimental control parameters and entering theoretical model coefficients. In this sense our model should be treated as a working
hypothesis. Although it provides a
qualitative insight to the intra-molecular vibrational dynamics, we are left with many questions unanswered
which must be perused in further work.
Understanding all its limitations, we nevertheless hope that our theory captures the essential elements of intra-molecular vibrational 
relaxation in
nano-systems.
Note that modern femtosecond spectroscopy methods (see e.g.,
\cite{BM98} -  \cite{BE02}, \cite{PM01} - \cite{SS04}) indeed demonstrate (in a qualitative
agreement with our consideration) remarkably different types of
behaviors (exponential decay and irregular damped oscillations) of
relatively close in energy initially excited states.
We believe that we are the first to explicitly address this issue.

\acknowledgments
Authors are indebted to Prof. W.H.Miller, and S.P.Novikov for stimulating discussions.
E.K. contribution 
to this research was supported by the National Science Foundation under Grant No PHY05-51164.

\end{document}